\documentclass[pra,twocolumn,showpacs]{revtex4}
\usepackage{graphicx}
\usepackage{amsmath}
\usepackage{times}

\begin{document}
\title{Topological Atom Laser}
\author{P. Blair Blakie$^{1,2}$, Robert J. Ballagh$^1$ and Charles W. Clark$^2$}
 \affiliation{$^1$Department of Physics, University of Otago,
 Dunedin, New Zealand}
  \affiliation{$^2$Electron and Optical Physics Division, National Institute of
  Standards and Technology, Gaithersburg, MD 20899-8410.} 
 \date{\today}

\begin{abstract}
We demonstrate how a topological atom laser can be realized by
output coupling a trapped vortex state with a Raman scattering
process.  We find a linearized analytic solution from which a
generalized resonance condition for Raman output coupling is
developed. Using numerical simulations of a two component
Gross-Pitaevskii equation in two and three dimensions the output
beam from a trapped central vortex state is analyzed for cases of
pulsed and continuous coupling where the vortex core is either
transverse or parallel to the direction of propagation. We show
how the parameters of the Raman light fields control the spatial
phase of the output beam.
\end{abstract}
\maketitle
\section{Introduction}
One of the most important technological developments to arise
to date from dilute gas Bose-Einstein condensation is the atom
laser \cite{Mewes1997a,Anderson1998a,Hagley1999a,
Bloch1999a,Martin1999a}. Analogously to its optical counter-part,
the atom laser produces highly coherent, directed matter waves,
and properties of this output beam, such as temporal and spatial
coherence, and beam divergence have been characterized
experimentally \cite{Kohl2001a,Trippenbach2000a,Coq2001a}. The
typical atomic laser configuration uses a confining magnetic
potential to act as a cavity for the laser mode, which is occupied
by a Bose Einstein condensate. Radio-frequency or optical Raman
transitions are then used to coherently transfer atoms into
untrapped hyperfine states which can propagate freely away from
the remaining trapped atoms \cite{Ballagh2000}. Usually the
condensate is taken to be in the ground motional state of its
confining potential, though it may possess thermal excitations
\cite{Choi2000a}.

In this paper we consider the matter wave output from a condensate
in a topologically excited state. To be definite, we take the
condensate to be in a central vortex state, though our formalism
is sufficiently general to apply to a broader range of topological
states including non-stationary states such as vortex arrays [e.g.,
see \cite{Madison2000a}]. We present a $T=0\,$K Gross-Pitaevskii
treatment of the topological atom laser dynamics, and give
numerical results that characterize the behavior over a wide
parameter regime, for two different scenarios: pulsed and
continuous scattering of the matter field with the output beam
direction transverse to the vortex core, and continuous scattering
in the direction parallel to the vortex core. We obtain a linearized
analytic solution for the output matter wave, and use
it to determine the resonance condition for the Raman process and
to characterize the phase properties of the atom laser.

\section{Formalism}
\subsection{Raman Coupling}
In this paper we consider a Raman output coupling mechanism of the
type experimentally demonstrated by Hagley \emph{et al.}
\cite{Hagley1999a}. In that scheme the absorption and emission of
optical photons from far detuned laser fields coherently transfers
atomic population between hyperfine states and imparts sufficient
momentum for the output coupled atoms to recoil relative to the
trapped condensate.

The ideal situation for an atom laser is to have only two
hyperfine states coupled: (1) a weak field seeking state which we
label $|1\rangle$ which is magnetically trapped by the confining
potential into the laser modes; (2) an untrapped state $|2\rangle$
in which atoms can freely propagate to form a directed output
matter wave.

The states $|1\rangle$ and $|2\rangle$ have an energy difference
$E_2 - E_1 = \hbar \omega_{21}$, and are coupled through an
intermediate state $|e\rangle$ by two laser fields, with frequency
$\omega_1$ and wavevector $\mathbf{k}_1$, and frequency $\omega_2$
and wavevector $\mathbf{k}_2$ respectively. The atom makes a
transition from state $|1\rangle$ to state $|2\rangle$ by
absorbing an $\omega_1$ photon, and emitting an $\omega_2$ photon,
thus transferring kinetic energy
\begin{equation}
\hbar \omega \equiv \hbar(\omega_1-\omega_2 )-\hbar \omega_{21}
\end{equation}
and momentum  $\hbar \mathbf{k} = \hbar(\mathbf{k}_1
-\mathbf{k}_2)$ to the atom's center of mass motion.  The laser
fields are assumed to be sufficiently far detuned from the
individual transitions that the effects of spontaneous emission
can be ignored, and the laser detunings can be represented by a
single value $\Delta$. We can then adiabatically eliminate the
intermediate state $|e\rangle$, and represent the laser field
coupling between states $|1\rangle$ and $|2\rangle$ by a two
photon Rabi frequency $V=\Omega_1 \Omega_2^*/2\Delta$, where
$\Omega_j$ is the single photon Rabi frequency associated with the
$\omega_j$ radiation field.

The condensate atoms will be scattered into the output state only
if the Raman energy transfer is approximately resonant i.e.
$\omega \approx \omega_k$ where $\hbar\omega_k = \hbar^2k^2/2m$
 is the recoil energy due to
the momentum transfer to the atoms. We note that the quadratic
dependence of the recoil energy on $k$ makes subsequent
Raman transitions into other Zeeman sublevels nonresonant. 
Thus provided $V$ is small
compared to $\omega_k$, any subsequent coupling  will be
negligible, and the system can be regarded as being effectively
two state.

\begin{figure}[!tbh]
{\centering \includegraphics[height=3.2in]{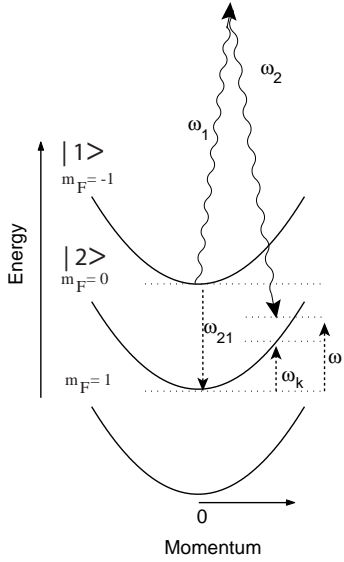}
\par}
\caption{\label{couplfig} Raman couplings between 
internal Zeeman sublevels and momentum states in an $F=1$
hyperfine manifold, appropriate for Rb$^{87}$ and Na$^{23}$.  The
frequencies of the two Raman lasers are $\omega_1$ and $\omega_2$
respectively, and the other quantities are defined in the text. }
\end{figure}

\subsection{Matter Wave Evolution}
The condensate atoms in internal state $|1\rangle$ have center of
mass wavefunction $\psi_1(\mathbf{r})$ and are trapped in the
cavity resonator mode by a harmonic potential $V_T(\mathbf{r})$.
Correspondingly, the atoms in state $|2\rangle$ have wavefunction 
$\psi_2(\mathbf{r})$.
The stationary modes of the trap (or resonator) satisfy the
Gross-Pitaevskii equation
\begin{equation}
\label{TIGPE} \hbar\mu\psi_1 =
-\frac{\hbar^2}{2m}\nabla^2\psi_1+V_T(\mathbf{r})\psi_1+w|\psi_1|^2\psi_1,
\end{equation}
where $\hbar\mu$ is the chemical potential and the collisional
interaction strength $w$ is given by $4\pi\hbar^2a_{11}N_0/m$ with
$a_{11}$ the s-wave scattering length for state $|1\rangle$ and
$N_0$ the number of condensate atoms.

For a topological atom laser we require that the trapped
condensate is in a topological state, whose properties will be
transferred to the output matter wave. For simplicity we consider
the case where the initial
condition for the condensate is a solution of Eq. (\ref{TIGPE}) that
 possesses a phase winding: in particular we will assume a central vortex
 state with a single quantum of circulation.

The time evolution of the trapped condensate and the outcoupled matter 
fields at $T=0\,$K is
given by the coupled Gross-Pitaevskii equations
\cite{Edwards1999a}
\begin{eqnarray}
i\hbar\frac{\partial\psi_1}{\partial
t}&=&-\frac{\hbar^2}{2m}\nabla^2\psi_1+V_T(\mathbf{r})\psi_1
+w(|\psi_1|^2+|\psi_2|^2)\psi_1\cr \cr &&+\frac{\hbar
V(t)^*}{2}e^{-i(\mathbf{k}\cdot\mathbf{r}-\omega
t)}\psi_2,\label{GPE1}\\  \cr i\hbar\frac{\partial\psi_2}{\partial
t}&=&-\frac{\hbar^2}{2m}\nabla^2\psi_2+w(|\psi_1|^2+|\psi_2|^2)\psi_2\cr
 \cr
  &&+\frac{\hbar
V(t)}{2}e^{i(\mathbf{k}\cdot\mathbf{r}-\omega
t)}\psi_1,\label{GPE2}
\end{eqnarray}
We have taken all scattering lengths between and within states to
be degenerate, and we note that the normalization condition for
the wavefunctions is $\int
d\mathbf{r}\,\left[|\psi_1|^2+|\psi_2|^2\right]=1$. In this paper
we solve Eqs. (\ref{GPE1}) and (\ref{GPE2}) numerically in two and
three spatial dimensions and also obtain an analytic solution for
the output wave.

\section{Linearized Analytic Solution}
 We will consider the regime where the depletion of the trapped condensate
is negligible on the time scale of the output coupling. This
condition will allow us to consider a perturbative solution for
the output beam amplitude, valid to first order in $V$. At this
order, we can ignore depletion of the trapped condensate, and
write
$\psi_1(\mathbf{r},t)=\sqrt{\rho(\mathbf{r})}\exp\left(iS_0(\mathbf{r})-i\mu
t\right)$ where $\rho(\mathbf{r})$($ \!=\!\!|\psi_1(\mathbf{r},
0)|^2$) and $S_0(\mathbf{r})$ are the density and phase profiles
of the initial trapped state respectively.  We will also assume that 
we can neglect the collisional interaction of atoms in state 
$|2\rangle$ on themselves, and on atoms in state $|1\rangle$ . This requires 
that the density of the scattered atoms is small compared to the trapped 
atom density in the region of overlap, which is consistent with the 
negligible depletion assumption.
Transforming $\psi_2$ to an interaction picture (to remove the
free particle-like motion) defined as
$\psi_2(\mathbf{r},t)=\bar\psi_2(\mathbf{r},t)\exp(i\mathbf{k}\cdot\mathbf{r}-i\omega
t)$, Eq. (\ref{GPE2}) can now be written in its linearized form as
\begin{eqnarray}
i\hbar\frac{\partial\bar\psi_2}{\partial t} &=&\left[
-\frac{\hbar^2}{2m}\nabla^2-\hbar\delta-i\hbar\mathbf{K}\cdot\nabla+
w\rho(\mathbf{r})\right]\bar\psi_2\cr &&+\frac{\hbar
V(t)}{2}\sqrt{\rho(\mathbf{r})}e^{iS_0(\mathbf{r})-i\mu
t},\label{linGPE}
\end{eqnarray}
where $\delta=\omega-\omega_k$ is the Raman detuning from free
particle resonance, $\mathbf{K}=\hbar\mathbf{k}/m$ is the recoil
velocity, and in accordance with the linearized treatment we have
ignored nonlinear terms involving  $\bar\psi_2$.

If the velocity spread in the trapped condensate is small compared to   
$K$ then we can ignore the effects of diffusion (about the 
central momentum $\hbar\mathbf{k}$) of the scattered 
wave packet by neglecting the Laplacian term in Eq. (\ref{linGPE}).
Noting that the local velocity of the trapped condensate at position ${\mathbf{R}}$ is 
\begin{equation}
\mathbf{v}(\mathbf{R})=\frac{\hbar \nabla _{\mathbf{R}}S_{0}(\mathbf{R})}{m},
\label{DefineLocalVelocity}
\end{equation}
this approximation is valid where $\mathbf{v}(\mathbf{R})^2\ll\mathbf{K}^2$. 
For typical experimental parameters this condition will be well satisfied everywhere
on the trapped condensate except close to the vortex core where the velocity 
field diverges. Making this approximation we obtain the
formal solution for the output coupled wave packet
\begin{eqnarray}
\bar\psi_2(\mathbf{r},t)=-\frac{i}{2}\int_0^t ds\,
e^{i\Theta(\mathbf{r},t,s)}V(s)\sqrt{\rho(\mathbf{r}+\mathbf{K}(s-t))},\label{formalsoln}
\end{eqnarray}
where
\begin{eqnarray}
\Theta(\mathbf{r},t,s)&=&S_0(\mathbf{r}+\mathbf{K}(s-t))-\mu
s+(t-s)\delta \cr
&&-\frac{w}{\hbar}\int_s^tds^\prime\,\rho(\mathbf{r}+\mathbf{K}(s^\prime-t)).\label{Theta}
\end{eqnarray}
In previous work we have presented a similar approximate solution
for Bragg scattering (where only a single internal state is
involved) and have verified that it provides a good representation
of the behavior of the Gross-Pitaevskii equation over a wide
parameter regime \cite{Blakie2001a}.

 The interpretation of  Eq. (\ref{formalsoln}) is similar to the Bragg case.  As the
scattered packet moves across the trapped condensate, the
amplitude $\psi_2$ at a given point (stationary in the frame
moving with velocity $\mathbf{ K}$ ) is built up from the sum of
contributions coupled in from the trapped condensate at successive
points along   the trajectory $\mathbf{R} = \mathbf{r}
+\mathbf{K}(s-t)$. The contribution coupled into $\psi_2$ at time
$s$ from position $\mathbf{R}$, has moved to position $\mathbf{r}$
at time $t$ and and has acquired a net phase of
$\Theta(\mathbf{r},t,s)$. If the phase term $\Theta$ varies
sufficiently slowly along a given trajectory, then the cumulative
contributions of matter scattered from the trapped condensate will
interfere constructively. Maximal scattering occurs when $\delta$
is chosen so that there is a point of stationary phase along the 
trajectory (i.e. $d\Theta(\mathbf{r},t,s)/{ds}=0$), which gives 
the generalized Raman resonance condition
\begin{eqnarray}
\delta \approx \left[\nabla_\mathbf{R}
S_0(\mathbf{R})\cdot\mathbf{K}-
\mu+w\rho(\mathbf{R})/\hbar\right]_{\mathbf{R}
=\mathbf{r}+\mathbf{K}(s-t)}. \label{GenRes}
\end{eqnarray}
It should be stressed that Eq. (\ref{GenRes}) is a local condition, which will apply
only at certain positions in the condensate. This leads to the important
property that the Raman scattering process can be spatially selective. The
resonance condition Eq. (\ref{GenRes}) differs from the Bragg resonance condition we
previously derived \cite{Edwards1999a}, and gives rise to the
following physical interpretation. 
Immediately after the Raman ejection process, an atom at $\mathbf{R}
$  will have velocity $\mathbf{K}+\mathbf{v}(\mathbf{R})$ [see Eq. 
(\ref{DefineLocalVelocity})]. The atom will
also have center of mass energy $\hbar (\omega +\mu ),$ arising from the
chemical potential released from the trapped condensate ($\hbar \mu $),
together with an additional energy $\hbar \omega $ gained from the radiation
field. This energy is divided between kinetic energy $m(\mathbf{K}+\mathbf{v}%
(\mathbf{R}))^{2}/2$ and potential energy  $w\rho (\mathbf{R)}$ arising from
interaction with the atoms remaining in the trapped condensate. Energy
conservation during the Raman process is therefore expressed as   
\begin{equation}
\hbar (\omega +\mu )=\frac{m}{2}(\mathbf{K}+\mathbf{v}(\mathbf{R}%
))^{2}+w\rho (\mathbf{R).}  \label{InterpResonance}
\end{equation}%
One can easily see that Eq. (\ref{InterpResonance}) is equivalent to Eq. (\ref{GenRes})
provided we neglect the term in $\mathbf{v}(\mathbf{R})^{2}$, consistent with
the validity condition $\mathbf{v}(\mathbf{R})^2\ll\mathbf{K}^2$ used in deriving 
the analytic solution.
It is also convenient below to interpret the term $\nabla S_{0}\cdot\mathbf{K}$  as
the Doppler shift due to the local velocity of the condensate. Two
particular cases of the resonance condition are worth examining:

\begin{itemize}
\item[(i)] For a non-interacting ground state (i.e. $S_{0}$ $=0,w=0$) the resonance
condition (\ref{GenRes}) is $\delta =-\mu .$ This is in contrast to the result for
Bragg scattering for this case, where resonance is at  $\delta =0,$ with the
difference due to the fact that the Raman process ejects the atom from the
trapped initial state,  so releasing the energy of that state.

 \item[(ii)] For an interacting ground state (i.e. $S_0=0$) the resonance condition is
$\delta= w\rho(\mathbf{R})/\hbar -\mu$. By density weighting the
contributions from all $\mathbf{R}$ across the condensate (assuming a Thomas-Fermi profile)
we find that the maximum amount of atoms will be scattered when
$\delta=-3\mu/7$. 
\end{itemize}

 \begin{figure}
{\centering \includegraphics[width=3.3in]{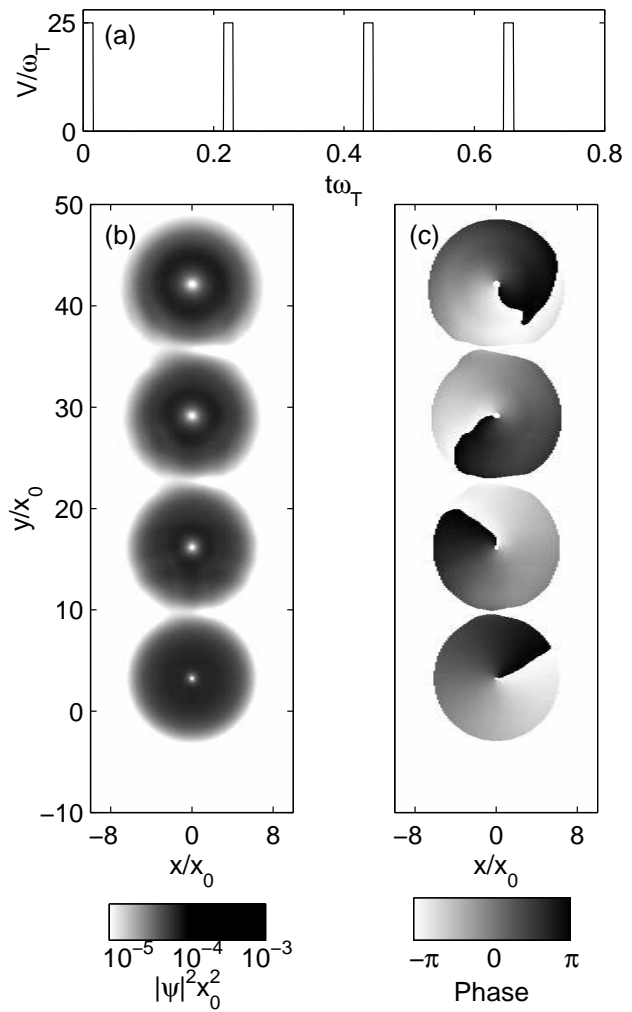}
\par}
\caption{\label{2DFig} Pulsed output matter wave from a 2D,
$m=-1$, central vortex state in a radially symmetric trap at
$t=0.71t_0$. (a) Temporal dependence of the Raman coupling $V(t)$.
(b) Output density profile. (c) Output phase profile. Raman output
coupler parameters: $\mathbf{k}=30\hat{\mathbf{k}}_y/x_0$, $\omega
=925\omega_T$, $\delta=25\omega_T$, and $V_{\rm
{max}}=25\omega_T$. Condensate parameters: $w=500w_0$ and
$\mu=9.2\omega_T$. Quantities are expressed in harmonic oscillator
units with $t_0=1/\omega_T$, $x_0=\sqrt{\hbar/2m\omega_T}$, and
$w_0=\hbar\omega_Tx_0^2$, where $\omega_T$ is the harmonic
trap frequency for the trapped condensate.}
\end{figure}

\section{Pulsed Output Coupling}
Here we consider a configuration of Raman fields which are pulsed
in time with a scattering direction (i.e. $\hat{\mathbf{K}}$)
perpendicular to the vortex core. With this choice, one axis of
the plane containing the characteristic spatial and phase
variations of the
 vortex is parallel to the direction of the beam propagation.
 For convenience we present 2D simulations of this geometry 
 with the Raman momentum difference vector along the $y$-direction, i.e.
$\mathbf{k}=k\hat{\mathbf{y}}$. The initial trapped state is
assumed to be  of the form $\psi_1(x,y)=R(r)e^{i\phi}$,
 where $r$ and $\phi$ are the
radial coordinate and polar angle respectively.

In Fig. \ref{2DFig} we show the output matter wave (i.e.
$\bar\psi_2$) after the application of four Raman pulses.
The length of each pulse $\tau _{p}$  is chosen to be sufficiently
short  that the atoms hardly move during the pulse, i.e. $\tau _{p}\ll R_{c}/K$
(where $R_{c}$ is the condensate radius), while the time between pulses is
sufficiently long that the scattered atoms can traverse the trapped
condensate before the next pulse is applied. The Raman intensity envelope 
$V(t)$ is shown in Fig. \ref{2DFig}(a), and the resulting output coupled density and
phase are shown in Figs. \ref{2DFig}(b) and (c). It is clear that this  procedure produces
copies of the trapped vortex state in the output state, including a $2\pi $ 
phase circulation, a result that we confirm analytically below.
We note that because $\bar\psi_2$ is defined 
in an interaction picture with respect to the free particle
center-of-mass motion, the rapid phase gradient due to the mean
wave packet velocity $\mathbf{K}$ has been removed, allowing the
phase information in Fig. \ref{2DFig}(c) to be more easily
discerned. For this reason all the phase plots we show in this
paper will be given for the interaction wavefunction.

The Fourier frequency width of each short pulse in our simulation
is $\sim210\,\omega_T$, which is much larger than the typical
frequencies associated with meanfield and Doppler shifts.
In this transient temporal regime, the resonance condition Eq. (\ref{GenRes}) 
is inapplicable, however the frequency spread is still sufficiently narrow 
that there is no significant coupling into other magnetic sublevels.
As long as the central frequency is close to resonance
($\omega\approx\omega_k$) this temporal-limited frequency spread
causes the Raman coupling to be resonant everywhere on the trapped
condensate facilitating the complete copying of the wavefunction
into the output component. The analytic solution Eq. (\ref{formalsoln}) 
contains the  vortex output result of Fig. \ref{2DFig} as
we can show by modeling an individual Raman pulse at time $t_{j}$ as a
delta function $V_{0}\delta (t-t_{j})$. For simplicity we shall choose $V_{0}
$ to be real, and then using Eq. (\ref{formalsoln}) we obtain for  $t>t_{j}$ 
(in the non-interaction picture)   
\begin{equation}
\psi _{2}(\mathbf{r},t)=-\frac{i}{2}V_{0}\sqrt{\rho (\mathbf{r}-\mathbf{R}%
_{j})}\exp (i\mathbf{k}\cdot \mathbf{r)}\exp (i\,\theta )  \label{psipulse}
\end{equation}
where $\mathbf{R}_{j}=\mathbf{K}(t-t_{j}),$ and 
\begin{eqnarray}
\theta &=&-\omega t+S_{0}(\mathbf{r}-\mathbf{R}_{j})-\mu t_{j}
+(t-t_{j})\delta\nonumber\\
&&-\frac{w}{\hbar }\int_{t_{j}}^{t}ds^{\prime}\,\rho (\mathbf{r}+\mathbf{K}(s^{\prime
}-t)).  \label{PhasePulse}
\end{eqnarray}
The density distribution of the output field in Eq. (\ref{psipulse}) is the
same as the initial vortex state,  but displaced by $\mathbf{R}_{j}$ (and
multiplied by $V_{0}/2$ ). The phase $\theta $ of the field can be written 
\begin{eqnarray}
\theta &=&\left[ S_{0}(\mathbf{r}-\mathbf{R}_{j})-\mu t_{j}\right] -\omega
t_{j}-\omega _{k}\left( t-t_{j}\right) \nonumber\\
&&-\frac{w}{\hbar }\int_{t_{j}}^{t}ds^{\prime}\,\rho
(\mathbf{r}+\mathbf{K}(s^{\prime }-t))  \label{PhaseInterp}
\end{eqnarray}
which has the following interpretation. The term in square brackets
represents the phase of the trapped vortex at time $t_{j}$ , which is
transferred to the output vortex along with the   phase of the Raman
coupling ($\omega t_{j}$)  at the time of the pulse.  In the absence of
collisional interactions the vortex would behave as a free particle of
momentum $\mathbf{k}$, and energy $\omega _{k}$ (we have neglected
diffusive kinetic energy) and its phase would increment as $\omega
_{k}\left( t-t_{j}\right) .$ The integral term in Eq. (\ref{PhaseInterp})
represents the correction to the free particle evolution during the time 
the output field is transiting the trapped vortex. Notice that the integrand
cuts off once $|\mathbf{r}+\mathbf{K}(s^{\prime }-t)|>R_{c},$ so that this
contribution to the phase becomes constant once the output vortex has
separated from  the initial vortex. The rotation of the  line of zero phase
between successive vortices in Fig. \ref{2DFig}(c) is thus described by the first 3
terms of Eq. (\ref{PhaseInterp}).

 \begin{figure}
{\centering \includegraphics[width=3.3in]{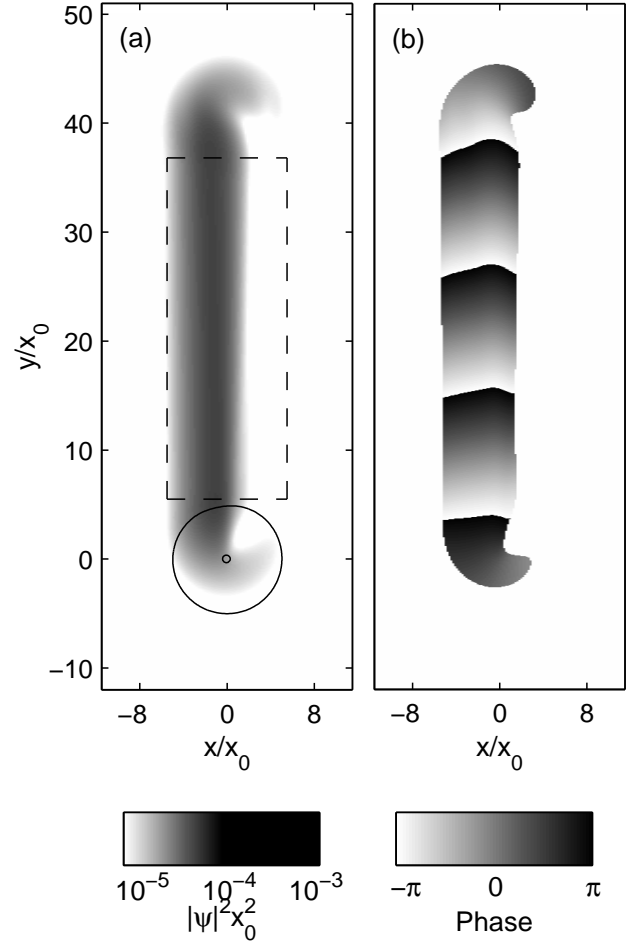}
\par}
\caption{\label{FigBeam} Continuously output coupled matter wave
from a 2D, $m=-1$, central vortex state of a radially symmetric trap
at $t=0.71 t_0$. with continuous Raman coupling. (a) Density
profile.  Dashed line indicates steady state beam region (see
text). The solid lines indicate density contours of the trapped 
vortex state.
(b) Phase profile. Raman parameters: $V=2$,
$\omega=925\omega_T$, $\mathbf{k}=30\hat{\mathbf{k}}_y/x_0$  and
$\delta=25\omega_T$. Condensate parameters: $w=500w_0$ and
$\mu=9.2\omega_T$. }
\end{figure}

\section{Continuous Output Coupling}
In Fig. \ref{FigBeam} we show the output matter wave following the
application of a long duration,  low intensity Raman pulse to a
vortex state - which we will refer to here as continuous output
coupling. This result displays typical behavior of the scattered
solution in the continuous regime: the matter wave is
preferentially scattered from a selected spatial region of the
trapped condensate and has an asymmetric density distribution
transverse to the direction along which it propagates. The shape
of the output density profile depends on the detuning $\delta$,
and by suitably adjusting this parameter, matter can be
selectively coupled out from either side of the trapped
vortex. Here the frequency width of the finite duration Raman
pulse is sufficiently narrow that the generalized resonance
condition (\ref{GenRes}) is applicable. From this resonance
condition, the spatial asymmetry of the output density profile can
be understood as arising from the Doppler term $\nabla_{
\mathbf{R}}S_0(\mathbf{R})\cdot\mathbf{K}$. A unit circulation
central vortex state with angular momentum number $m =\pm1$ has a
phase profile of the form $S_0=\pm\phi$, where $\phi$ is the
azimuthal angle. The Doppler effect gives a local shift of the resonant
frequency;  on the side of the vortex where the current flow runs
parallel to the scattering direction the resonant frequency is
shifted upward, while on the other side the flow is anti-parallel
and the frequency is shifted downward. By choosing $\delta$  above
or below the central resonance value, it is possible to
selectively scatter from either side of the vortex.

To investigate the properties of the continuously output matter
wave in more detail we consider the steady
state region of the output coupled matter wave. 
This region, which we indicate pictorially in Fig. \ref{FigBeam}(a), 
is where the output beam has constant transverse density profile, that
we shall refer to as the \emph{steady state beam profile}. From
the formal solution of Eq. (\ref{formalsoln}) it can be shown that
for two points $\mathbf{r}$ and $\mathbf{r}+\alpha\mathbf{K}$,
both within the steady
state beam region, then the values of the wavefunction at these
points are related by
\begin{equation}
\bar\psi_2(\mathbf{r}+\alpha\mathbf{K},t)=\bar\psi_2(\mathbf{r},t)
e^{i(\delta+\mu)\alpha}.\label{BeamTranslation}
\end{equation}
This result indicates that the steady-state  velocity of the output matter
wave is $\mathbf{K+V}_{0},$ where we define    
\begin{equation}
\mathbf{V}_{0}=\frac{\hbar }{m}\frac{\delta +\mu }{K}\mathbf{\hat{K}.}
\label{OutputVelocity}
\end{equation}%
Remarkably, this is uniform across the transverse profile, a result we can
understand using Eq. (\ref{GenRes}), which shows that  at the position where the Raman
transfer takes place, the trapped condensate has a  local velocity component
parallel to $\mathbf{K}$ given by 
\begin{equation}
\mathbf{v}(\mathbf{R})_{\Vert }=\frac{\hbar }{m}\frac{\left( \delta +\mu
-w\rho (\mathbf{R})\right) }{K}\mathbf{\hat{K}}.
\label{ParallelLocalVelocity}
\end{equation}%
The velocity of the atom (in direction $\hat{\mathbf{K}}$) immediately after the
Raman transfer is therefore  $\mathbf{K+v}(\mathbf{R})_{\Vert }.$  We showed
in section III that this local ejection velocity gives a kinetic energy
consistent with an atom residing in a  potential of $w\rho (\mathbf{R),}$
(due to the trapped condensate atoms). Once the atom has moved away from the
condensate, this potential energy is converted to kinetic energy, consistent
with an atom of velocity $\mathbf{K+V}_{0}.$ We note that  the transverse
components of the local ejection velocity  cancel out in the final 
scattered wave, due to the symmetry of the vortex velocity distribution. For
a general straight-line trajectory across the vortex, there are two positions
of local resonance with the same density $\rho (\mathbf{R),}$ but with
opposite transverse local velocities.  
It is also worth emphasizing that the control available over the output
matter wave velocity is in contrast with the case of optical lasers, where
the group velocity of the light cannot be altered without changing the
properties of medium in which it propagates.

\begin{figure}
{\centering \includegraphics[width=2.8in]{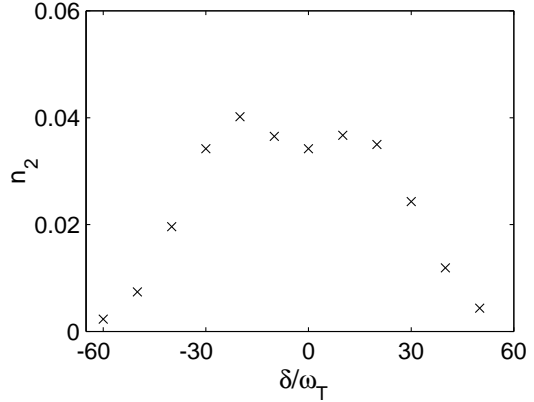}
\par}
\caption{\label{phasepropsfig} Population scattered into an output matter
wave from a vortex after excitation by a continuous Raman pulse of duration
$t=0.71t_0$. Raman parameters:
 $\mathbf{k}=30\hat{\mathbf{k}}_y/x_0$ and
$V=2\omega_T$. Condensate parameters: $w=500w_0$ and
$\mu=9.2\omega_T$.}
\end{figure}

We have numerically verified that Eq. (\ref{BeamTranslation})
is in good agreement with numerical simulations of
the Gross-Pitaevskii Equation over a wide regime.  We note that
changing the Raman detuning $\delta$ to control the phase gradient
of the output coupled beam also affects the efficiency at which the
atoms are scattered, and to characterize this we show the
scattered population, i.e. $n_2=\int
d\mathbf{r}|\bar\psi_2(\mathbf{r},t)|^2$ in Fig.
\ref{phasepropsfig}. This result shows that in this case the frequency 
width of the Raman coupling   (from $\delta \sim -30\omega_T$ to $30\omega_T$) 
is sufficient to allow the  output velocity  to be  changed 
over a range of $1.0x_0 \omega_T$  without
significant attenuation of the output matter wave. In these
results the Doppler effect is the dominant broadening mechanism,
so that the approximate width of the Raman transition,
$\Delta\omega$, will be
\begin{equation}
\Delta\omega=\frac{k\Delta p}{m},\label{Dopfreqwidth}
\end{equation}
where $\Delta p$ is the momentum width of the condensate in the
scattering direction. For the results in Fig. \ref{phasepropsfig}
Eq. (\ref{Dopfreqwidth}) gives the full width at half maximum 
$\Delta\omega=60.8\omega_T$, where
$\Delta p$ is numerically evaluated from the initial vortex
eigenstate (the initial state for $\psi_1$).

Additionally, it should be noted that the scattered population
($n_2$) is a convenient experimental observable for measuring the
linear response of the condensate to the perturbing Raman
potential. This technique for probing the condensate is similar to
Bragg spectroscopy, which was first used on condensates by Stenger
\emph{et al.} \cite{stenger1999b}, except that for the Raman case
the condensate is scattered into a different internal state. This
technique has several advantages over Bragg spectroscopy for
conducting high precision measurements of condensate properties
which we will explore elsewhere.

\begin{figure}
{\centering \includegraphics[width=3in]{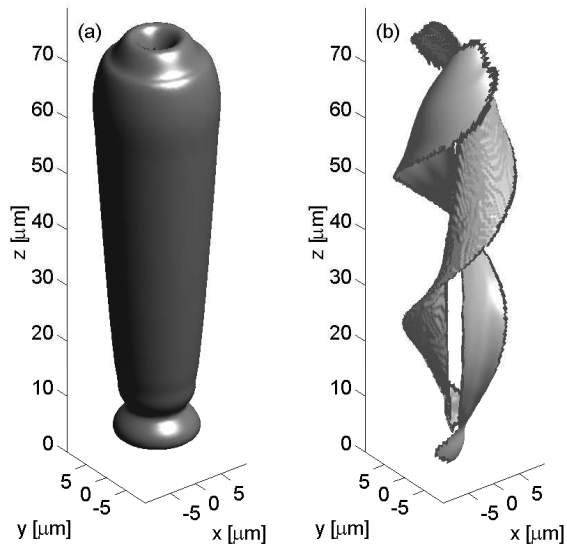}
\par}
\caption{\label{CylFig1} Output matter wave at $t=1.4$ms from a
$2.2\times10^5$ atom $^{23}$Na condensate in a $m=1$ central
vortex state with $\mu=1.15$kHz in a spherically symmetric trap of
frequency $80$Hz. Raman output coupler corresponds to $589$nm
laser fields at $120^\circ$ with $\delta=0$ and a two-photon
Raman frequency of $V=2\pi\times480$s$^{-1}$. (a)
$3.4\times10^{12}$atom/cm$^3$ density iso-surface of the output
matter wave. (b) $0$ and $\pi$ phase iso-surfaces of the output
matter wave.}
\end{figure}

Finally we consider continuous output coupling of a central vortex
in a direction parallel to the vortex core. Because this
arrangement has the superfluid current flow perpendicular to
$\mathbf{K}$, the Doppler term and hence the output coupling is
not spatially selective\footnote{There is a residual resonance
shift due to the inhomogeneous meanfield, but we have found that
typically this does not give rise to significant spatial
selectivity if the condensate chemical potential is of order or
smaller in size that the Doppler width (see Eq.
(\ref{Dopfreqwidth})).}. This means that the vortex profile will
be reproduced as the transverse profile of the atom laser output,
so that the atom laser will have a similar intensity and phase
profile to that of a TEM$^*_{01}$ mode of an optical laser,
commonly known as the donut mode \cite{Siegman}.

In Figs. \ref{CylFig1} (a) and (b) we show phase and density
iso-surfaces of a vortex output-coupled in the manner described
above, for realistic experimental parameters. The noticeable beam
divergence in the stable beam region, which extends from about
$z=10\mu$m to $z=60\mu$m, is enhanced by the centrifugal forces 
of the rotating condensate beyond that expected simply from 
repulsive wave packet spreading

The helicity of the phase iso-surface arises from the addition of
the phase gradient along the propagation direction (due to the
Raman output coupling) to the vortex-like phase profile the matter
wave has in the transverse direction. The phase gradient of the
matter wave and hence the phase helicity is controlled by Raman
detuning, as given in Eq. (\ref{BeamTranslation}). We have
numerically verified that for a case with the same parameters as in Fig.
\ref{CylFig1}, except using  $\delta\sim-\mu$, the helicity
disappears as the phase iso-surfaces are approximately vertical.
Also for $\delta <-\mu$ the helicity changes sign
corresponding to the phase iso-surfaces spiraling in the opposite
sense.

The helical structure of the phase could be investigated by an
interference experiment, such as superimposing the topological
atom laser beam with a co-propagating plane phase atom laser.

\section{Conclusion}
In this paper we have introduced the idea of a topological atom
laser and considered its behavior using numerical and analytic
approaches over a broad parameter regime. We have developed a
linearized solution for the output coupled matter wave from which
we have determined a spatially dependent resonance condition for
Raman scattering and characterized how the output beam phase
properties relate to the Raman detuning and the condensate
chemical potential.

\section{Acknowledgments}
This work was supported by the US Office of Naval Research, 
the Advanced Research and Development Activity. PBB and RJB 
acknowledge support from the Marsden Fund of New Zealand 
under contracts PVT902 and PVT202.


\end{document}